**PAPER • OPEN ACCESS**

# Singularities and Phenomenological aspects of Asymptotic Safe Gravity

To cite this article: V Zarikas and G Kofinas 2018 *J. Phys.: Conf. Ser.* **1051** 012028

View the article online for updates and enhancements.

# Singularities and Phenomenological aspects of Asymptotic Safe Gravity

V Zarikas[1,2] and G Kofinas[3]

[1]Nazarbayev University, School of Engineering, Astana, Kazakhstan
[2]University of Applied Sciences of Central Greece, Lamia, Greece
[3]Research Group of Geometry, Dynamical Systems and Cosmology, University of the Aegean, Samos, Greece
E-mail: vasileios.zarikas@nu.edu.kz; gkofinas@aegean.gr

**Abstract.** Asymptotic Safety (AS) Program for quantum gravity keeps the same fields and symmetries with General Relativity and studies the associated gravitational action as a fundamental part of the complete theory at the nonperturbative level with the help of functional renormalization group (RG) techniques. An important phenomenological task that can test the new point of view of AS approach is the discovery of RG improved cosmologies and black holes. In this work, we analyze the properties of recently found non-singular spherically symmetric and non-singular cosmological solutions. Furthermore, we derive a novel consistent set of modified Einstein field equations, in the spirit of AS, which respects the Bianchi identities. This new set of equations completes previously published modified Einstein equations which arise by adding appropriate covariant kinetic terms to the action.

## 1. Introduction
The singularity theorems of Hawking and Penrose erased the hope that spacetime singularities in classical General Relativity are just a consequence of the high degree of symmetries assumed in the collapse or in the energy momentum tensor. Now we know that the appearance of spacetime singularities is a proof of the limit of validity of General Relativity. An important property of any serious attempt to quantize gravity is the ability to resolve the breakdown of physics in these singular regions of spacetime. Loop Quantum Gravity and spin foam models offer ideas to solve this problem since they assume that the spacetime is intrinsically discrete. Concerning models on continuous spacetimes, one interesting proposal is that of Asymptotic Safety Program. Asymptotic Safety (AS) framework for quantum gravity keeps the same fields and symmetries with General Relativity and studies the associated gravitational action as a fundamental part of the complete theory at the nonperturbative level (with the help of functional renormalization group (RG) techniques) [1-60].

The key idea behind the AS scenario was first proposed by Weinberg. The central requirement is the existence of a non-Gaussian fixed point (NGFP) of the RG flow for gravity. If this NGFP exists, this defines the behavior of the theory at the UV. Furthermore, in this regime all measured quantities are free from nonphysical divergences. Asymptotic safe gravity does not emerge from a direct quantization of General Relativity. The Einstein-Hilbert action is a bare action that corresponds to a non-trivial fixed point of the RG flow and is a prediction assuming asymptotic safety. The AS framework justifies someone to work with the gravitational effective average action which includes only the effect of the quantum fluctuations with momenta $p^2 > k^2$. Thus the effective average action represents an approximate description of physics at the momentum scale $p^2 \approx k^2$. In this way, it is possible to develop, following various different paths, a phenomenological study of asymptotic safety.

The scope of this work is to investigate the structure of spacetime at high energies and in particular both in the vicinity of the center of spherically symmetric solutions and in the big bang regime of cosmological solutions. Both these type of solutions are generated from quantum modified Einstein equations according to the AS framework. The present paper has two targets. First, we analyze the properties of recently found non-singular spherically symmetric and cosmological solutions. Secondly, we present a novel improved version of modified Einstein equations according to the AS idea. The proposed new equations are compatible with a new alternative and natural scheme for energy conservation. The way we will produce these new modified Einstein equation, valid for short distance interiors of spherically symmetric configurations and early times cosmologies, follows. AS framework uses a system of truncated RG flow equations that contains two running couplings, the gravitational constant $G(k)$ and the (positive) cosmological constant $\Lambda(k)$. Near the non-Gaussian UV fixed point the coupling $G$ approaches zero, while on the other hand, the coupling $\Lambda$ behaves as the square of the momenta. Due to this scaling of $G$, $\Lambda$ in the UV regime, we work on the belief that in the Big bang regime and in the center of black hole the $\Lambda$ term plays the dominant role. Thus, we also study the modified Einstein vacuum equations where only $\Lambda(k)$ appears. More precisely, the result of the existence of the non-Gaussian fixed point at the UV is utilized. This allows us to adopt the scaling behaviour of $\Lambda$ as $\Lambda \sim k^2$.

With a running cosmological constant as the source, novel uniquely defined covariant gravitational equations that modify minimally the Einstein gravity are derived. They are constructed so that the Bianchi identities are satisfied by construction. We first have to find the appropriate covariant kinetic terms that support an arbitrary source term $\Lambda(k)$ without any symmetry assumption and construct modified Einstein equations which respect the Bianchi identities. If we add a spacetime-dependent cosmological constant $\Lambda(x)$ in the 4-dimensional vacuum Einstein equations so that $G_{\mu\nu} = -\Lambda g_{\mu\nu} + 8\pi G T_{\mu\nu}$, obviously, this equation does not satisfy the Bianchi identities. Thus, we will add an energy-momentum tensor $\theta_{\mu\nu}$ to support $\Lambda(x)$, so that the Bianchi identities are satisfied. This formalism has the advance that allows the term $\Lambda(x)$ to be kept arbitrary (later we will specify the function $\Lambda(x)$).

The outline of the paper is as follows: In the second section of the present work a review of recently found solutions from quantum modified equations is presented. In the third section a novel scheme of quantum modified equations is developed with a different energy conservation scheme. The novel scheme of energy conservation includes all the matter energy-momentum tensor and the varying couplings $G$, $\Lambda$. Finally, in the last section various conclusions are discussed.

## 2. Review of recently found solutions

In this section we will try to provide an insight into recently developed solutions from quantum modified Einstein equations according to the AS framework, [61-62]. These new modified Einstein equations describe how a classical spacetime may appear in the presence of a quantum vacuum. The quantum vacuum is modeled through a cosmological constant term which is energy dependent according to the AS program.

Vacuum solutions where the cosmological constant is solely a source are of great importance. A motivation for this is that the vacuum birth of our universe from a vacuum fluctuation is an interesting proposal in various quantum gravity models. The same also holds for the central regions of black holes where we expect spacetime to be associated with antigravity properties. Thus, the vacuum modified Einstein equations have been constructed with the scope to be valid in the UV regime. In this transplanckian regime the cosmological constant plays the dominant role according to AS. This reason why in [61-62] only kinetic terms for $\Lambda(k)$ are included. A spacetime-dependent cosmological constant $\Lambda(k)$ has to be introduced in the 4-dimensional vacuum Einstein equations so that $G_{\mu\nu} = -\Lambda g_{\mu\nu}$. However, this equation is inconsistent since the cosmological constant term is not any more constant. Thus, it does not satisfy the Bianchi identities. Therefore, it is necessary to add an energy-momentum tensor $\theta_{\mu\nu}$ to support $\Lambda(k)$, with the demand that Bianchi identities are satisfied. The new modified Einstein equations become

$$G_{\mu\nu} = -\Lambda g_{\mu\nu} + \theta_{\mu\nu}. \tag{1}$$

This equivalently gives

$$R_{\mu\nu} = \Lambda g_{\mu\nu} + \theta_{\mu\nu} - \frac{1}{2}\theta g_{\mu\nu}, \tag{2}$$

where $\theta = \theta^{\mu}{}_{\mu}$. Now the Bianchi identities are

$$\theta_{\mu\nu}{}^{;\mu} = \Lambda_{;\nu}, \tag{3}$$

where ; denotes covariant differentiation. For $\Lambda$ constant, the quantity $\theta_{\mu\nu}$ should vanish since in this case the Bianchi identities are trivially satisfied and there is no need to add anything.

Assuming that the tensor $\theta_{\mu\nu}$ can be constructed by $\psi$ and its first and second derivatives, it is possible to express it as $\theta_{\mu\nu} = A\psi_{;\mu}\psi_{;\nu} + Bg_{\mu\nu}\psi^{;\rho}\psi_{;\rho} + \tilde{N}\psi_{;\mu;\nu} + Eg_{\mu\nu}\Box\psi + Fg_{\mu\nu}$, where $\Box\psi = \psi_{;\mu}{}^{;\mu}$. Note that $\psi$ is defined as $\Lambda = \bar{\Lambda}e^{\psi}$ where $\bar{\Lambda}$ is an arbitrary constant reference value. It is possible to solve the system of equations (3) and find finally that $\theta_{\mu\nu}$ is given by $\theta_{\mu\nu} = -\frac{1}{2}\psi_{;\mu}\psi_{;\nu} - \frac{1}{4}g_{\mu\nu}\psi^{;\rho}\psi_{;\rho} + \psi_{;\mu;\nu} - g_{\mu\nu}\Box\psi$. From this expression we obtain the self-consistent modified Einstein equations in terms of $\Lambda$. Note that $\psi$ does not have its own equation of motion since we are working in the AS framework.

Now one crucial question is how we can consistently include an extra matter energy-momentum tensor $T_{\mu\nu}$ in the vacuum field equations. Let us write

$$G_{\mu\nu} = -\bar{\Lambda}e^{\psi}g_{\mu\nu} - \frac{1}{2}\psi_{;\mu}\psi_{;\nu} - \frac{1}{4}g_{\mu\nu}\psi^{;\rho}\psi_{;\rho} + \psi_{;\mu;\nu} - g_{\mu\nu}\Box\psi + 8\pi G T_{\mu\nu}. \tag{4}$$

Note that the gravitational constant $G$ is in general also spacetime dependent. In order to satisfy the Bianchi identities we have to take into account that the various $\psi$-kinetic terms in (4) are by construction covariantly conserved. Thus in order to satisfy the Bianchi identities we have to conserve energy-momentum in combination with $G$, i.e.

$$\left(GT_{\mu\nu}\right)^{;\nu} = 0. \tag{5}$$

If one insists on the exact energy-momentum conservation, then either extra covariant terms constructed out of $G$ would be necessary to be added in (4) or the approach presented in the next section is needed. In case we allow kinetic terms for both $\Lambda$ and $G$, then this has the serious drawback that at the end the modified Einstein equations are very hard to be solved. Such an approach with starting from an action that generates equations of motion with quantum corrections of $G$ and $\Lambda$ can be found in [4]. However, in this approach the full $\Lambda$ and $G$-dependent terms, which make the equations consistent, have not been found in general. Thus, our philosophy behind the construction of both versions of modified Einstein equations ala AS, is that we keep only the kinetic terms of $\Lambda$ for both simplicity/tractability and vacuum domination at UV. A remark for equations (4) is that it is not clear if they arise from an action or not.

Finally, the RG flow provides the exact energy dependent law for $\Lambda(k)$, $G(k)$, and the system of equations (4), (5) is a consistent system of equations valid near the NGFP in the presence of matter. The "cosmological constant" term scales at high energies as the asymptotic safety scenario suggests, i.e. $\Lambda(k) \propto k^2$. The generation of solutions needs also the connection of the momentum scale $k$ with some length scale.

In [61], several families of novel spherically symmetric solutions have been derived in the context of AS. The analysis of all the solutions provides evidence for the spacetime structure close to the centre of spherical symmetry. The signature of a spherically symmetric metric close to this centre is not known, and this is the reason for considering both options, not only the one with the "conventional" signs for the temporal-radial metric components $(-+++)$, but also the alternative one with signature $(+-++)$. It is well known that the second signature appears in the interior Schwarzschild black hole for the same spherical coordinates.

There are two categories of physically significant solutions, i.e. solutions which improve or resolve the singularity problem. The first category is represented by one solution with the "conventional" signature that originates from $r=0$ and is associated with a curvature singularity there. The interesting feature is that no timelike geodesic can reach the origin since the geodesic of an infalling massive particle is reflected at some finite distance. The explanation of this behavior is that there is a repulsive force caused by the cosmological constant along with the other kinetic terms. Note that this feature is not met by a standard cosmological constant. However, the massless particles reach the origin in finite affine parameter.

The second category concerns several solutions of both types of signatures. These solutions are not extended to include the centre of spherical symmetry and range from a finite radius outwards. All their curvature invariants at this minimum radius are also finite. Thus, they can also be considered as non-singular physically accepted geometries.

In [62], general families of new cosmological solutions have been derived in the context of Asymptotic Safe Gravity at high energies close to the NGFP. The derived vacuum solutions were found from the same set of vacuum modified Einstein equations as in [61]. In addition an extension of the vacuum modified Einstein equations is presented in [62], capable to describe also matter. The new equations use a certain scheme for energy conservations described from Eq. (5). Thus, the work [62] can describe two different cosmic eras. First, a speculated vacuum birth from a quantum vacuum state and second a radiation-matter evolution at high energies.

In the first such cosmic era the only source is an energy-dependent cosmological constant that scales at the UV as the AS framework proposes near the NGFP. The importance of the derived physically interesting vacuum solutions is the fact that they describe an inflationary expansion with completely removed initial singularity for the scale factor, the energy density and the curvature invariants. The end of inflation happens naturally since in the AS scenario, as the energy scale becomes lower, $\Lambda(k)$ becomes insignificant and standard decelerating cosmology arises.

In the second cosmic era, an energy exchange due to Eq. (5) is realized between the matter and the varying gravitational constant $G(k) \propto k^{-2}$. An interesting discovery is the appearance of negative non-equilibrium pressure beyond the thermodynamic one. This can be interpreted as either a particle production or a mechanism of bulk viscosity. In both alternatives, there are general solutions which are non-singular and inflationary. The same holds for any spatial topology. A suspiring finding is that in the case of bulk viscosity the relation between the energy scale and the time is implied by the theory of AS. Finally, perhaps the most important feature of the matter solutions is that they suggest either particle production with entropy generation or bulk viscosity with entropy production and reheating.

**3. New Field equations with a varying cosmological constant**
Here, in this section we provide an alternative version of quantum modified Einstein equations following an alternative approach for encapsulating Bianchi identities. We also add a spacetime-dependent cosmological constant $\Lambda(k)$ in the 4-dimensional vacuum Einstein equations so that $G_{\mu\nu} = -\Lambda g_{\mu\nu} + 8\pi G T_{\mu\nu}$. In the framework of asymptotic safety, $\Lambda(k)$ is supposed to be determined along the RG flow. Nevertheless, we want to construct a general formulation valid for an arbitrary $\Lambda(k)$ which consequently will be able to carry any $\Lambda(k)$ specified by RG equations. As we have already explained, the previous equation is inconsistent since it does not satisfy the Bianchi identities. Therefore an energy-momentum tensor $\theta_{\mu\nu}$ will be introduced in order to support $\Lambda(k)$, so that the Bianchi $-4\pi G(Tg_{\mu\nu} - 2T_{\mu\nu})$ identities are satisfied. In the new equations to be derived here, the key point is that the matter energy-momentum tensor will participate in the construction and satisfaction of the Bianchi identities.

The gravitational equations of motion are
$$G_{\mu\nu} = -\Lambda g_{\mu\nu} + \theta_{\mu\nu} + 8\pi G T_{\mu\nu} \tag{6}$$
or equivalently
$$R_{\mu\nu} = \Lambda g_{\mu\nu} + \theta_{\mu\nu} - \frac{1}{2}\theta g_{\mu\nu} - 4\pi G T g_{\mu\nu} + 8\pi G T_{\mu\nu}. \tag{7}$$

It is $R = 4\Lambda - \theta - 8\pi G T$, where $\theta = \theta^{\mu}{}_{\mu}$, $T = T^{\mu}{}_{\mu}$. The Bianchi identities take the form
$$\theta_{\mu\nu}{}^{;\mu} = \Lambda_{;\nu} - 8\pi(G T_{\mu\nu}{}^{;\mu} + T_{\mu\nu}G^{;\mu}), \tag{8}$$
where ";" denotes as usually covariant differentiation with respect to the Christoffel connection of $g_{\mu\nu}$. We introduce for convenience the quantity $\psi(x)$ by
$$\Lambda = \overline{\Lambda}e^{\psi}, \tag{9}$$
where $\overline{\Lambda}$ is an arbitrary constant reference value.

Let us construct such a tensor $\theta_{\mu\nu}$ from $\psi$ and its first and second derivatives. Therefore, we work with the ansatz
$$\theta_{\mu\nu} = A(\psi)\psi_{;\mu}\psi_{;\nu} + B(\psi)g_{\mu\nu}\psi^{;\rho}\psi_{;\rho} + \tilde{N}(\psi)\psi_{;\mu;\nu} + E(\psi)g_{\mu\nu}\Box\psi + F(\psi)g_{\mu\nu}, \tag{10}$$
where $\Box\psi = \psi_{;\mu}{}^{;\mu}$. Then,
$$\theta = (A+4B)\psi^{;\mu}\psi_{;\mu} + (C+4E)\Box\psi + 4F, \tag{11}$$
$$R_{\mu\nu} = A\psi_{;\mu}\psi_{;\nu} - \left(\frac{A}{2}+B\right)g_{\mu\nu}\psi^{;\rho}\psi_{;\rho} + \tilde{N}\psi_{;\mu;\nu} - \left(\frac{C}{2}+E\right)g_{\mu\nu}\Box\psi + (\overline{\Lambda}e^{\psi}-F)g_{\mu\nu}. \tag{12}$$

Since $\Box(\psi_{;\nu}) = (\Box\psi)_{;\nu} + R_{\nu\sigma}\psi^{;\sigma}$, it is from (10)
$$\theta_{\mu\nu}{}^{;\mu} = (A'+B')\psi^{;\mu}\psi_{;\mu}\psi_{;\nu} + (A+E')\Box\psi\psi_{;\nu} + (C'+A+2B)\psi^{;\mu}\psi_{;\mu;\nu} + \\ + (C+E)(\Box\psi)_{;\nu} + F'\psi_{;\nu} + C R_{\mu\nu}\psi^{;\mu}, \tag{13}$$
where a prime denotes differentiation with respect to $\psi$. Due to Eq. (12), Eq. (13) can be written as
$$\theta_{\mu\nu}{}^{;\mu} = \left(A'+B'+\frac{1}{2}AC-BC\right)\psi^{;\mu}\psi_{;\mu}\psi_{;\nu} + \left(E'+A-CE-\frac{1}{2}C^2\right)\Box\psi\psi_{;\nu} + \\ + (C'+A+2B+C^2)\psi^{;\mu}\psi_{;\mu;\nu} + (C+E)(\Box\psi)_{;\nu} + (F'-CF+C\overline{\Lambda}e^{\psi}-4\pi GCT)\psi_{;\nu} + \\ + 8\pi GCT_{\mu\nu}\psi^{;\mu}. \tag{14}$$

The consistency condition Eq. (8) becomes
$$\left(A'+B'+\frac{1}{2}AC-BC\right)\psi^{;\mu}\psi_{;\mu}\psi_{;\nu} + \left(E'+A-CE-\frac{1}{2}C^2\right)\Box\psi\psi_{;\mu} + (C'+A+2B+C^2)\psi^{;\mu}\psi_{;\mu;\nu} + \\ + (C+E)(\Box\psi)_{;\nu} + \left[F'-CF+(C-1)\overline{\Lambda}e^{\psi}-4\pi GCT\right]\psi_{;\nu} = -8\pi(GC\psi^{;\mu}+G^{;\mu})T_{\mu\nu} - 8\pi G T_{\mu\nu}. \tag{15}$$

One option to proceed would be to embody the term $-8\pi GC\psi^{;\mu}T_{\mu\nu}$ of the right-hand side of Eq. (15) with the last term of the left-hand side being proportional to $g_{\mu\nu}\psi^{;\mu}$. Since Eq. (15) must be satisfied for any $\psi$, the various prefactors should vanish separately and the prefactor of $\psi^{;\mu}$ would give $T_{\mu\nu} \propto g_{\mu\nu}$ with a function as a proportionality factor (the vanishing of the rest terms on the right-hand side would give an equation of motion for the matter). Such an energy-momentum tensor is of a very special form, not of particular interest, so we quit this option. The other alternative is to vanish the prefactors of the various kinetic terms on the left-hand side as they stand. The right-hand side gives the equation of motion for the matter. Therefore, we obtain the following system of equations

$$A' + B' + \frac{1}{2}AC - BC = 0, \quad E' + A - CE - \frac{1}{2}C^2 = 0$$
$$C' + A + 2B + C^2 = 0, \quad C + E = 0$$
$$F' - CF + (C-1)\overline{\Lambda}e^\psi - 4\pi GCT = 0$$
$$GT_{\mu\nu}{}^{;\mu} + (GC\psi^{;\mu} + G^{;\mu})T_{\mu\nu} = 0.$$
(16)

It can be seen that the first equation is redundant and the previous system gets the form

$$A + B = -\frac{3}{4}C^2, \quad E = -C, \quad C' = \frac{1}{2}C^2 + A$$
$$F' - CF + (C-1)\overline{\Lambda}e^\psi - 4\pi GCT = 0$$
$$GT_{\mu\nu}{}^{;\mu} + (GC\psi^{;\mu} + G^{;\mu})T_{\mu\nu} = 0.$$
(17)

The differential equation for $F$ is linear, so its solution is

$$F = e^{\int C}\left\{\sigma + \int\left[(1-C)\overline{\Lambda}e^\psi + 4\pi GCT\right]e^{-\int C}\right\},$$
(18)

where $\sigma$ is integration constant. Since for $\psi$ constant, $\theta_{\mu\nu}$ must vanish, it arises from Eq. (10) that for $\psi$ constant the factor $F$ must vanish. From the form of the previous solution for $F$, for this to happen for any constant $\psi$, it needs to be $C=1, T=0$ and furthermore $\sigma=0$. Thus, although in the system of Eqs. (17) there are four equations containing the five unknowns $A, B, C, E, F$, the above requirement gives a unique solution. The system of Eqs. (17) is written equivalently as

$$A = -\frac{1}{2}, \quad B = -\frac{1}{4}, \quad C = 1, \quad E = -1, \quad F = 0$$
(19)

$$T = 0$$
(20)

$$T_{\mu\nu}{}^{;\mu} + \left(\psi^{;\mu} + \frac{G^{;\mu}}{G}\right)T_{\mu\nu} = 0.$$
(21)

For any given $\Lambda$, $G$ the non-conservation equation (21) defines the equation of motion of the matter. Our final generic result is equation

$$G_{\mu\nu} = -\overline{\Lambda}e^\psi g_{\mu\nu} - \frac{1}{2}\psi_{;\mu}\psi_{;\nu} - \frac{1}{4}g_{\mu\nu}\psi^{;\rho}\psi_{;\rho} + \psi_{;\mu;\nu} - g_{\mu\nu}\psi + 8\pi GT_{\mu\nu},$$
(22)

where $T_{\mu\nu}$ satisfies the two equations (20), (21). The vanishing of the trace of the energy-momentum tensor, $T=0$, is a surprising outcome of the study, since it constraints the possible equations of state in an interesting way. For the case of a perfect fluid, this equation results to a radiation-dominated Universe. This is a result which is compatible with what we expect near the NGFP taking into account that the system of equations could have suggested something totally unacceptable as possible equation of state near the NGFP. It is also noticeable that although the new energy conservation scheme in the present work is totally different than the one used in [62], leading to different system of equations, still the kinetic terms of the cosmological constant coupling are the same, thus Eq. (22) is the same. Finally, the non-conservation equation (21) expresses an energy exchange between the matter fields and the couplings $G(k)$, $\Lambda(k)$.

As in the previous section with the vacuum solutions, we also here assume a spatially homogeneous and isotropic metric for the cosmic spacetime of the form Eq. (3). Since the external fields $\Lambda(k)$, $G(k)$ carry the same symmetries, they will be of the form $\Lambda(t)$, $G(t)$. We consider a diagonal energy-momentum tensor $T^\mu_\nu$, so we take as matter content a non-perfect fluid with energy density $\rho$, thermodynamic pressure $p$ and a possible non-equilibrium part $\pi$ [54-57]. Due to the working symmetries of isotropy and homogeneity shear viscosity and energy fluxes are disregarded. The energy momentum tensor is

$$T^{\mu\nu} = \rho u^\mu u^\nu + (p+\pi)(g^{\mu\nu} + u^\mu u^\nu) \qquad (23)$$

with $u^\mu$ the fluid 4-velocity. The extra pressure $\pi$ can either be associated to a pressure due to particle production/destruction or to a bulk viscous pressure and could be important during the transition phase that connects the quantum vacuum stage of the universe to the subsequent era with non-zero matter density. We will consider in the following a very early high-energy universe with the assumption of AS running of the couplings.

The two independent components of (22) are

$$3\left(H^2 + \frac{\kappa}{a^2}\right) = \bar{\Lambda}e^\psi - 3H\frac{\dot\psi}{n} - \frac{3\dot\psi^2}{4n^2} + 8\pi G\rho \qquad (24)$$

$$\frac{2}{n}\dot H + 3H^2 + \frac{\kappa}{a^2} = \bar{\Lambda}e^\psi - 2H\frac{\dot\psi}{n} - \frac{\dot\psi^2}{4n^2} - \frac{1}{n}\left(\frac{\dot\psi}{n}\right)^\cdot + 8\pi GP, \qquad (25)$$

where the total effective pressure is $P = p + \pi$ and $n$ denotes the lapse function.

In the NGFP regime of AS it is $G(k) \propto k^{-2}$, $\Lambda(k) \propto k^2$, therefore Eq. (21) implies the exact conservation $T_{\mu\nu}{}^{;\mu} = 0$. Thus, although energy can be exchanged between $G(k)$ and $\Lambda(k)$, the matter energy momentum tensor is separately conserved something that it is correct and expected in our semiclassical approach at the tree level! Alternatively, if someone demands $T_{\mu\nu}{}^{;\mu} = 0$ (again motivated by the correct tree level behavior) it can be proven generally that $G(k)\Lambda(k) = const$, which is exactly the expected behavior at the NGFP! Thus, a consistent picture arises, considering the derived modified Einstein equations associated with the energy conservation scheme of Eq. (8).

For cosmology this conservation equation gives the standard expression

$$\dot\rho + 3nH(\rho + P) = 0. \qquad (26)$$

The system of equations (24)-(25) is satisfied by construction for any $\psi(t)$. However, the effective pressure $P$ is constrained to be $P = \frac{1}{3}\rho$. This is different than the result obtained in [62] where the opposite sign appered in the equation if state. Indeed, here differentiating (23) with respect to $t$ and using (26) to substitute $\dot\rho$ and also (24) itself we find in the UV regime

$$\left(1 + \frac{\dot\psi}{2nH}\right)\left[\frac{2}{n}\dot H + 3H^2 + \frac{\kappa}{a^2} - \bar{\Lambda}e^\psi + 2H\frac{\dot\psi}{n} + \frac{\dot\psi^2}{4n^2}G(k) + \frac{1}{n}\left(\frac{\dot\psi}{n}\right)^\cdot + 8\pi GP\right] = 4\pi G\frac{\dot\psi}{nH}\left(P - \frac{1}{3}\rho\right). \qquad (27)$$

Therefore, from Eq. (25) it arises that $P = \frac{1}{3}\rho$. For vanishing $\pi$, we obtain $p = \frac{1}{3}\rho$, which is a radiation equation of state, quite natural in the early universe. Our result is equation (24) with $\rho \propto a^{-4}$ and $\psi$ is expressed in terms of $k$ (and the other cosmological variables) as explained above.

Far from the UV regime, the evolution of $G(k)$, $\Lambda(k)$ is not the simple power laws and Eq. (21) gets the more general form

$$\dot{\rho} + 3nH(\rho + P) + \rho\left(\frac{\dot{G}}{G} + \dot{\psi}\right) = 0 . \tag{28}$$

This equation expresses a non-trivial energy exchange between the energy density and the time-dependent couplings and has to be solved together with Eq. (24), (25).

**4. Conclusions**
In this paper, we present the non singular properties of recently found spherical and cosmological solutions of quantum modified Einstein equations according to AS framework. The general solutions contain a variety of smoothing behaviors and several explicit non-singular solutions. The solutions solve the uniquely defined modified Einstein equations which arise by adding appropriate covariant kinetic terms to the equations of motion in order to ensure the satisfaction of the Bianchi identities. We assume that the cosmological constant kinetic terms dominate near NGFP, something that it is expected in the AS. The importance of the presented solutions lies on the fact that they may describe the small distance behavior of quantum corrected spacetimes. In the case of spherical solutions, we claim that these solution may describe the non-singular geometry at the center of black holes, while in the case of cosmology the non-singular solutions most probably represent the geometry of Big bang after a vacuum birth according to asymptotically safe gravity. The cosmological solutions could also be utilized towards other directions that explore high energy corrections to the expansion rate for explaining baryogenesis [63] or dark energy [64].
Although the modified Einstein equations ala AS are uniquely defined, the way we satisfy energy conservation has three alternatives. In this paper we present for the first time the remaining two alternatives of energy conservation we can have. These new two schemes of energy conservation accompany consistently the modified Einstein equations. A surprising fact is that both these schemes are associated with constrained form of equations of state that are compatible with the UV regime. The fact that the possible energy momentum tensors are constrained from the type of energy conservation is desirable and sensible. It is sensible since the form of matter and its interactions are determining which type of energy exchange is allowed and which is not. Furthermore, one of these schemes predicts a radiation equation of state in the early universe.

**Acknowledgements**
Vasileios Zarikas acknowledges the hospitality and support of Nazarbayev University (social policy SPG).